\documentclass{article}

\usepackage[utf8]{inputenc}
\usepackage{aas_macros}
\usepackage[english]{babel}
\usepackage[section]{placeins}

\usepackage[letterpaper,top=2cm,bottom=2cm,left=3cm,right=3cm,marginparwidth=1.75cm]{geometry}

\usepackage{amsmath}
\usepackage{graphicx}
\usepackage[colorlinks=true, allcolors=blue]{hyperref}
\usepackage{authblk}
\usepackage{multicol}
\usepackage[T1]{fontenc}
\DeclareUnicodeCharacter{0300}{\`{}}
\DeclareUnicodeCharacter{0301}{\'{}} 
\usepackage{natbib}

\usepackage{longtable}

\date{accepted in PSS on January 27, 2025}

\begin{document}
\title{The refractory-to-ice ratio in comet 67P: implications on the composition of the comet-forming region of the protoplanetary disk}

\author[1]{Raphael Marschall}
\author[2]{Alessandro Morbidelli}
\author[3]{Yves Marrocchi}

\affil[1]{Observatoire de la Côte d'Azur, Laboratoire J.-L. Lagrange, CNRS, CS 34229, 06304 Nice Cedex 4, France}
\affil[2]{Coll\`{e}ge de France, Centre National de la Recherche Scientifique, Universit\'{e} Paris Sciences et Lettres, Sorbonne Universit\'{e}, 75014 Paris, France}
\affil[3]{Universit\'{e} de Lorraine, CNRS, CRPG, UMR 7358, 54000 Nancy, France   }

\maketitle

\begin{abstract}
Comets, asteroids, and other small bodies are thought to be remnants of the original planetesimal population of the Solar System.
As such, their physical, chemical, and isotopic properties hold crucial details on how and where they formed and how they evolved.
Yet, placing precise constraints on the formation region of these bodies has been challenging.
Data from spacecraft missions have a particularly high potential of addressing the question of the origin of the visited bodies.
ESA's Rosetta mission to comet 67P/Churyumov-Gerasimenko returned data from the comet for two years on its journey around the Sun.
This extensive data set has revolutionised our view on comets and still holds unsolved problems.
   
Here, we aim to determine comet 67P's bulk elemental composition from Rosetta data, including its refractory-to-ice ratio.
We use these results to constrain the temperature in the protoplanetary disk where comets formed and, using a disk model, the formation location.

We use the Rosetta/ROSINA (Rosetta Orbiter Spectrometer for Ion and Neutral Analysis) measurement of the volatile/ice composition and the Rosetta/COSIMA (COmetary Secondary Ion Mass Analyzer) measurements of the refractory composition of comet 67P.
These measurements are combined using a Monte Carlo method.
The refractory-to-ice ratio is a free parameter that is constrained a posteriori.

Using only the composition, we constrain the refractory-to-ice ratio to $0.5<\chi<1.7$, and derive the bulk elemental abundances for 67P of H, C, N, O, Na, Mg, Al, S, K, Ar, Ca, Cr, Mn, Fe, Kr, and Xe.
We find the noble gas xenon in near solar elemental abundance in comet 67P.
Krypton is slightly depleted, while argon is heavily depleted.
Comet 67P is enriched in all three noble gases by up to 2.5 orders of magnitude compared to CI chondrites.
We show this is consistent with a formation region between 25 and 35~au in a protoplanetary disk region with temperatures between 30 and 40 K and with the trapping of dust for a long time in rings of the protoplanetary disk.

\end{abstract}

%\begin{multicols}{2}

%%%%%%%%%%%%%%%%%%%%%%%%%%%%%%%%%%%%%%%%%%%%%%%%%%%%%%%%%%%%%%%%%%%%%%%%%%%%%%%%%%%%%%%%%%%%
%%%%%%%%%%%%%%%%%%%%%%%%%%%%%%%%%%%%%%%%%%%%%%%%%%%%%%%%%%%%%%%%%%%%%%%%%%%%%%%%%%%%%%%%%%%%
%%%%%%%%%%%%%%%%%%%%%%%%%%%%%%%%%%%%%%%%%%%%%%%%%%%%%%%%%%%%%%%%%%%%%%%%%%%%%%%%%%%%%%%%%%%%
%%%%%%%%%%%%%%%%%%%%%%%%%%%%%%%%%%%%%%%%%%%%%%%%%%%%%%%%%%%%%%%%%%%%%%%%%%%%%%%%%%%%%%%%%%%%
%%%%%%%%%%%%%%%%%%%%%%%%%%%%%%%%%%%%%%%%%%%%%%%%%%%%%%%%%%%%%%%%%%%%%%%%%%%%%%%%%%%%%%%%%%%%
%%%%%%%%%%%%%%%%%%%%%%%%%%%%%%%%%%%%%%%%%%%%%%%%%%%%%%%%%%%%%%%%%%%%%%%%%%%%%%%%%%%%%%%%%%%%
\section{Introduction}\label{sec:introduction}
It is commonly argued that studying small Solar System bodies (asteroids, comets, etc.) directly informs planet formation theories because they are remnants of the original planetesimal population that triggered Solar System formation.
Comets are considered among the most pristine objects in our Solar System because their physical and chemical properties are most easily explained by a lack of internal processing.
For example, comets have very low densities \citep[$\sim500$~kg~m$^{-3}$;][]{Groussin2019} and contain significant amounts of highly-volatile ices \citep[incl. CO and CO$_2$;][]{Eberhardt1987A&A,Gasc2017MNRAS}.
Although most comets have typical diameters between 1-5~km \citep{Snodgrass2011MNRAS}, the mentioned properties appear to hold even for large comets \citep[e.g.,][]{Biver1997EM&P, Kelley2022ApJL, Spencer2020, Brown2013ApJL}, suggesting that these bodies experienced minimal heating or other planetary processes.

This view that comets are, to a large extent, primordial is consistent with our current understanding of their dynamical past.
They have presumably formed beyond Neptune in a massive primordial disk of precursory bodies, commonly referred to as planetesimals or cometesimals \citep[e.g.][for a review of the dynamics in the early Solar System]{Nesvorny2018ARA&A}.
When Neptune migrated through that disk, these bodies were scattered into the current day Kuiper Belt -- in particular, the so-called Scattered Disk --  where they have remained until recently \citep[e.g.][]{Nesvorny2016ApJL, Nesvorny2021ApJL}.
The Scattered Disk is widely considered as the main source reservoir of Jupiter family comets \citep[JFC;][]{Duncan2004come, Duncan2008SSRv, Dones2015SSRv}.
From the Scattered Disk, comets are injected into the inner Solar System by the giant planets. 
Once close enough to the Sun, they exhibit activity driven by the sublimation of ices (in particular H$_2$O).

Because JFCs can have very low perihelia ($\sim 1$~au), they are also the most easily accessible comet population for interplanetary spacecraft.
Over the past three decades, they have thus become the target of multiple international missions, including 19P/Borrelly \citep[Deep Space 1,][]{Rayman2002AcAau}, 81P/Wild 2 \citep[Stardust,][]{Reichhardt1995Natur}, 9P/Tempel 1 \citep[Stardust, and Deep Impact,][]{AHearn2005SSRv}, 103P/Hartley 2 \citep[Deep Impact/EPOXI,][]{AHearn2011}, and most recently 67P/Churyumov-Gerasimeko \citep[Rosetta,][]{Taylor2017RSPTA}.

Rosetta was the first mission to go into orbit around a comet.
Two years at 67P/Churyumov-Gerasimenko (hereafter 67P) gave us an unprecedented up-close look at the physical and chemical properties as well as the evolution of 67P along its orbit from $3.7$~au on the inbound orbit to perihelion at $1.25$~au and back out to $3.8$~au.
Rosetta has revolutionised our understanding of comets \citep[see][for a review]{Keller2020SSRv}.
Yet many questions remain open.
In particular, the hope that we can determine how comets formed has not been fulfilled.
Proponents of conflicting formation scenarios have claimed Rosetta's observations confirm their theories \citep[see][]{Weissman2020SSRv}.

One main objective, finding the formation location of comet 67P, has not yet been addressed, and it is one of our goals.
In this work, we combine measurements from two analytical instruments from the Rosetta space mission: i) ``Rosetta Orbiter Spectrometer for Ion and Neutral Analysis" (ROSINA) and ii) ``COmetary Secondary Ion Mass Analyzer" (COSIMA) instruments to deduce the bulk chemical composition of 67P.
We will show that the noble gas xenon is present in solar abundance, which strongly suggests that it has been accreted in solid form, while argon was accreted out of the gas phase.
Krypton being slightly depleted is suggestive that the comet formed close, but slightly inward, of the krypton sublimation line.
This implies the formation of comet 67P and JFCs, more generally, at roughly $25-35$ au in the protoplanetary disk.
This constraint is solely driven by composition but is consistent with dynamical models on the origin and evolution of Jupiter family comets.

\section{Deriving the bulk composition of comet 67P}
The bulk elemental abundance of 67P was not measured directly.
Instead, two instruments measured two different components, which we call here the ice (anything measured as a gas in the coma) and refractory (measured on solid dust particles) components.
The ice component was measured by the Rosetta Orbiter Spectrometer for Ion and Neutral Analysis \citep[ROSINA;][]{Balsiger2007SSRv}.
The refractory component was measured on dust particles collected by the COmetary Secondary Ion Mass Anaylzer \citep[COSIMA;][]{Kissel2007SSRv}.

To get the bulk elemental composition of 67P, we need to combine these two components.
To do so and to properly track the errors and uncertainties, we model the data from both ROSINA and COSIMA, assuming that the respective measured values represent the median of a log-normal distribution and the respective errors are 1-$\sigma$ errors.
Taking one million random draws of each element with the respective distribution gives a good match to the data, as shown in the appendix (Fig.~\ref{fig:compare2ROSINAandCOSIMA}).

\subsection{The refractory component from COSIMA}
The COSIMA instrument collected dust particles in the coma of 67P.
These particles were, amongst other things, analysed with a high-resolution time-of-flight secondary ion mass spectrometer.
The elemental abundances of the cometary dust that we are using here were presented in Table~3 of \cite{Bardyn2017MNRAS}.
The Nitrogen to carbon ratio, N/C, was measured by \cite{Fray2017MNRAS}.

Here, we use all the elemental abundances with respect to iron, Fe.
But, because N/Fe was not directly measured the 'minus'-error became very large when deriving it from the measurements of the N/C, C/Si, and Si/Fe ratios.
Because this large error leads to a negative lower bound for the N/Fe ratio, which is non-physical, we have chosen to use the logarithm of the 'positive' error also for the 'negative' error.
This, in principle, can overestimate the abundance of Nitrogen, but given that Nitrogen is almost two orders of magnitude depleted with respect to the major elements of hydrogen, carbon, and oxygen, we believe this assumption is justified and will not affect our results in a significant way.

Another assumption of the COSIMA measurements is that the hydrogen-to-carbon atomic ratio is $1.04 \pm 0.16$ \citep[][]{Isnard2019A&A}.
This is justified by the fact that the macro-molecules that were detected in
67Ps particles contain more H than the Insoluble Organic Matter (IOM) measured in chondritic meteorites \citep[][]{Fray2016Natur}.
The highest H/C in IOM that was measured is 0.8 \citep{Alexander2007GeCoA}.
We thus adopt this same assumption as \cite{Isnard2019A&A} and \cite{Bardyn2017MNRAS}.
Additionally, COSIMA found that the refractory component of 67P is made up by roughly one half in organics and the other half in minerals.
All used measurements from \cite{Bardyn2017MNRAS} and \cite{Fray2017MNRAS}, as well as the log-normal parameters used to model the measurements, are listed in Table~\ref{tab:cosima-measurements}.

To zeroth order, the refractory component of 67P consists of equal parts of hydrogen, carbon, and oxygen (left panel of Fig.~\ref{fig:compare2ROSINAandCOSIMA}).
The next most abundant element, silicon, is a factor of four less abundant than the three major elements.
All other species are at least 1.5 orders of magnitude less abundant than the major elements.
Potassium is the least abundant of the measured elements and is depleted by more than three orders of magnitude compared to hydrogen.

\subsection{The ice component from ROSINA}
The composition of the gases in the coma of 67P was measured by the ROSINA instrument.
ROSINA consisted of three subsystems: the Double-Focusing Mass Spectrometer (DFMS), the Reflectron-type Time-Of-Flight mass spectrometer (RTOF), and the Comet Pressure Sensor (COPS).
\cite{Rubin2019MNRAS} determined the molecular abundances of 45 major and minor volatile species with respect to water.
We use these values to derive the elemental abundances with respect to water (right panel of Fig.~\ref{fig:compare2ROSINAandCOSIMA}).
Table~\ref{tab:rosina-measurements} lists the ROSINA measurements and the log-normal parameters used to model the measurements.

Compared to the refractory component, the volatiles are depleted in carbon. 
Hydrogen and oxygen are still the major elements, while sulphur and nitrogen are both less abundant by two orders of magnitude, similar to the refractory component.
The three noble gases argon, krypton, and xenon are five to seven orders of magnitude less abundant than hydrogen.

\subsection{Constraining the refractory-to-ice mass ratio}
To combine the refractory and volatile/ice components and arrive at the bulk elemental composition of 67P, we are left with one unknown: the refractory-to-ice mass ratio, $\chi$.
There is considerable debate on the value of $\chi$ for 67P, or any comet for that matter \citep[see review by][]{Choukroun2020SSRv}.
Values from 0.1 to 10 and higher can be found in the literature.
The large discrepancies in these values result primarily from a lack of knowledge about how much dust falls back to the surface and to what extent that dust has lost its ice component while on ballistic trajectories around the comet.

\cite{Rubin2019MNRAS} derived bulk elemental abundances for hydrogen, carbon, oxygen, nitrogen, and silicon, assuming $\chi$ equal to one and three.
Here, we take a different approach and start by being agnostic about the value of $\chi$ and thus initially leave it as a free parameter.
We will see later that doing that will allow us to independently constrain $\chi$ using only arguments of the elemental composition.
Additionally, we derive the bulk elemental abundances for 12 additional elements (Na, Mg, Al, S, K, Ar, Ca, Cr, Mn, Fe, Kr, Xe), which crucially also includes the noble gases argon, krypton, and xenon.

Thus, for each value of $\chi$, we draw a random composition for the refractory and volatile/ice components, assuming the log-normal distribution described above.
Each component is mixed with the other at the respective $\chi$, from which we can calculate the resulting bulk elemental composition.
By performing one million such random draws, we can calculate the median elemental composition and the associated 1-$\sigma$ errors.

\begin{figure}
	\includegraphics[width=\textwidth]{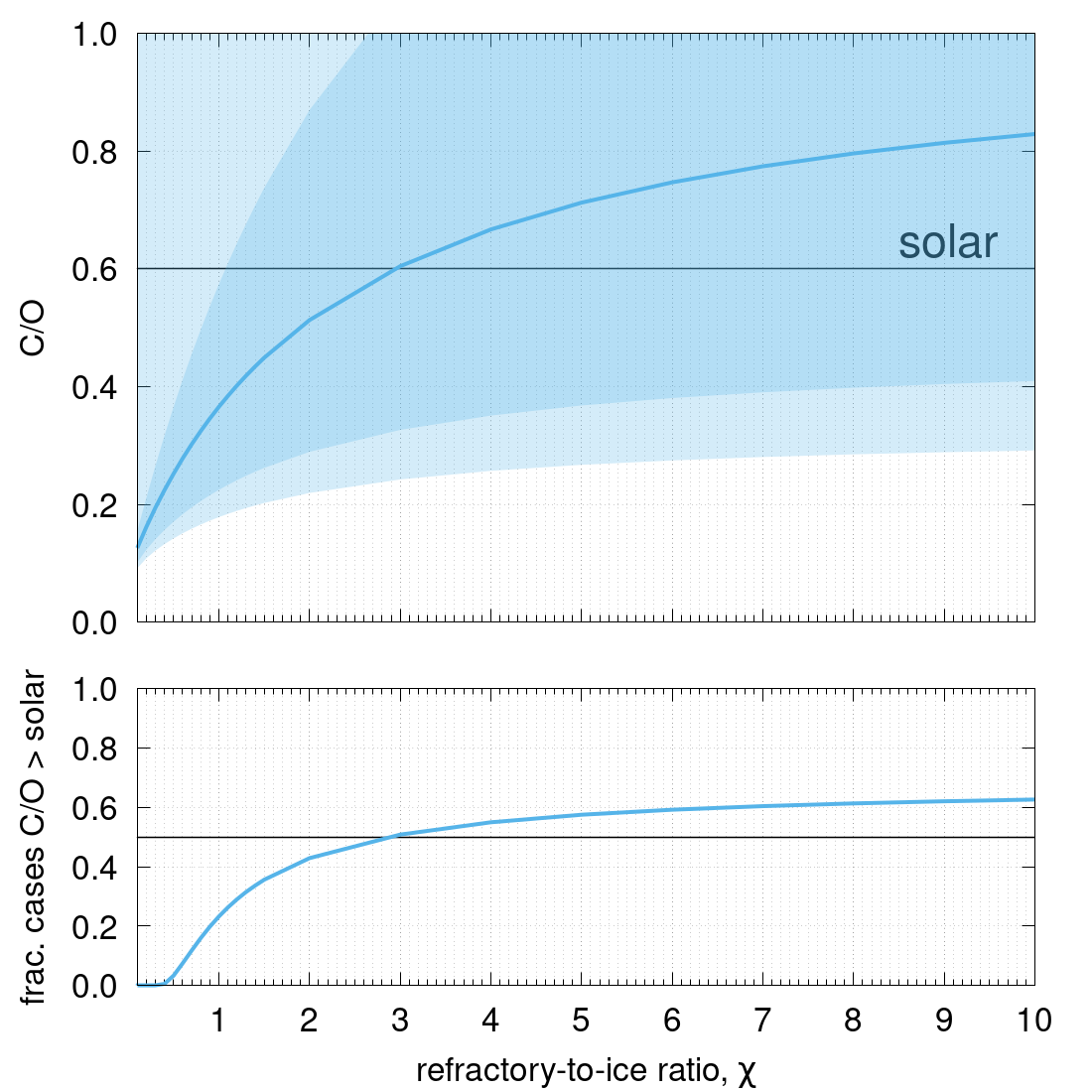}
	\caption{The top panel shows the bulk carbon-to-oxygen atomic ratio of comet 67P as a function of the refractory-to-ice mass ratio, $\chi$. The dark blue shaded area encompasses 50\% of the cases, and the light blue $\pm 1 \sigma$. The solar value of 0.594 is taken from \cite{Asplund2021A&A}. The bottom panel shows the fraction of cases with a super-solar C/O ratio.}
	\label{fig:C2O} 
\end{figure}

Figure~\ref{fig:C2O} shows the resulting carbon-to-oxygen ratio as a function of $\chi$.
The dark blue shaded area encompasses 50\% of the cases, and the light blue $\pm 1 \sigma$.
The error bars are large and become larger as $\chi$ increases because of the large uncertainties of the C and O abundances of the refractory component.

It would be reasonable to assume that the C/O ratio could be as high as the solar value \citep[][]{Asplund2021A&A} but should not be super-solar.
In the median case, this is true up to $\chi=3$.
The bottom panel of Fig.~\ref{fig:C2O} show the fraction of cases with a super-solar C/O ratio.
It shows that even at high $\chi$ there are still roughly $40\%$ of cases where 67P would have a sub-solar C/O ratio.
This might be acceptable to some, but we would posit that if the condition of having a sub-solar C/O ratio is strongly weighted, that would clearly point in the direction of smaller $\chi$.
For example at a $\chi=1$ almost $80\%$ of cases would be sub-solar.
Nevertheless, the C/O ratio, though pointing towards small $\chi$, is not a strong constraint on $\chi$.

\begin{figure}
	\includegraphics[width=\textwidth]{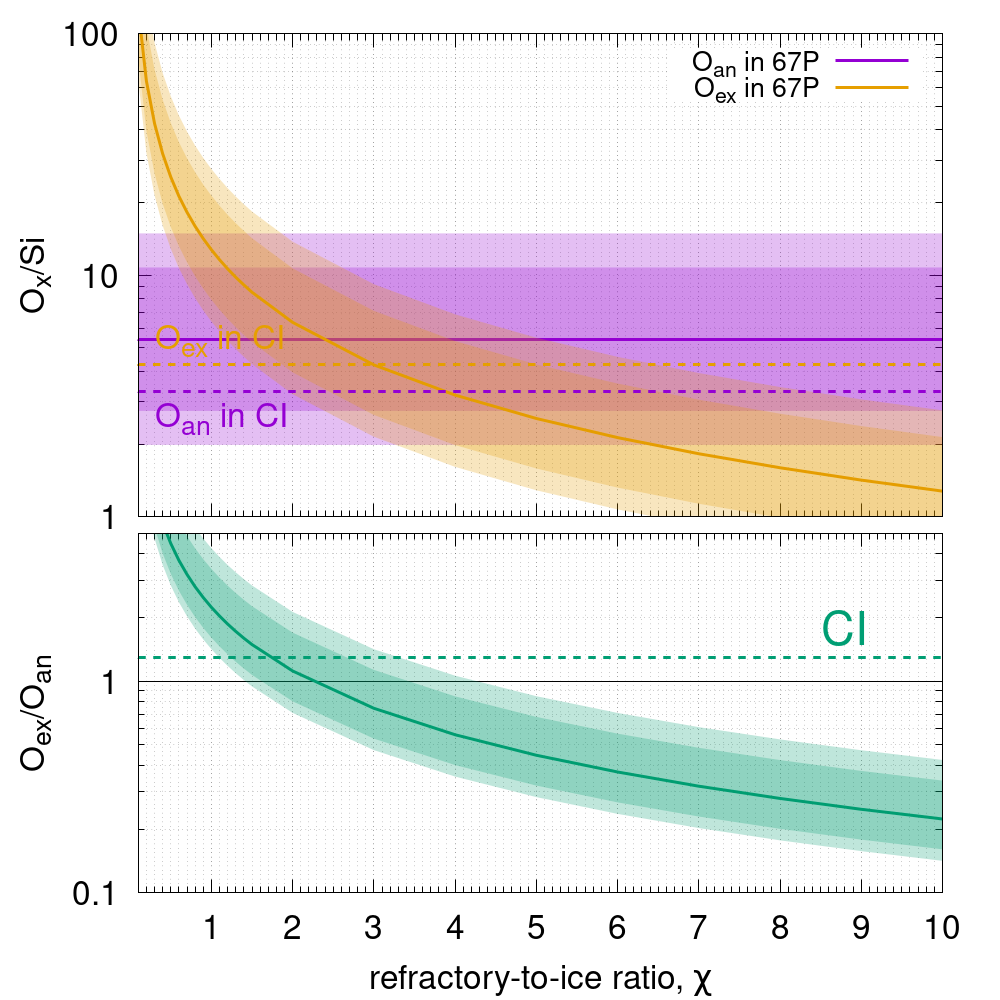}
	\caption{The top panel shows the oxygen-to-silicon ratio for oxygen that is in water (subscript 'ex') and that which is in anhydrous form (subscript 'an'). The values for CI chondrites are taken from \cite{Alexander2019GeCoA}. The bottom panel shows the ratio of the two oxygen components. Note that the bottom panel is not simply the ratio of the lines in the top panel. O$_{ex}$ and O$_{an}$ are not uncorrelated, and therefore, their ratio and respective errors need to be computed for each random draw. The darker shaded areas contain 50\% of the cases while the lighter areas contain $\pm 1 \sigma$}
	\label{fig:oxygen} 
\end{figure}

Therefore, we will have a look next at the oxygen composition as a function of $\chi$.
This is particularly interesting because we can split the oxygen into a part that is stored in water (O$_{ex}$) and the one that is stored in anhydrous material (O$_{an}$).
This is analogous to what has been done for meteorites \citep[][]{Alexander2019GeCoA}, e.g., the primitive CI chondrites.
Figure~\ref{fig:oxygen} shows the oxygen composition of 67P and its comparison to CI chondrites.
The anhydrous oxygen composition compared to silicon is independent of $\chi$ because it is simply the composition of the refractory component.
But the amount of oxygen in water (with respect to silicon) decreases with increasing $\chi$ because the fraction of water decreases with increasing $\chi$.

In the median case, if $\chi=1.7$, the ratio between the abundance of oxygen in water and anhydrous minerals is the same as in CI chondrites (Fig.~\ref{fig:oxygen}).
This case already seems unusual, as we might expect a comet to have more oxygen in water than a carbonaceous chondrite.
The comet is non-chondritic if $\chi<1.1$.
Nevertheless, we consider the $\chi=1.7$ case ("chondritic" case) the upper limit.
Remember that for $\chi<3$, the median C/O ratio is also always sub-solar and therefore, for $\chi<1.7$, most cases will result in a sub-solar C/O ratio.

To summarise this part:
When starting with an agnostic stance on what the refractory-to-ice mass ratio, $\chi$, for 67P should be, from the elemental composition alone constrains $\chi<1.7$.
This is lower than what was found very early in the mission \citep[$\chi=4\pm2$;][]{Rotundi2015Sci} but in line with measurements from the coma of 67P \citep[][]{Biver2019A&A,Marschall2020b, Laeuter2020, Combi2020}.
Initially, there was a significant discrepancy between the estimates for the gas production rates derived from remote sensing (e.g., the MIRO and VIRTIS instrument) and in-situ measurements (ROSINA).
Later studies resolved this, and now the different estimates are in line with each other \citep{Biver2019A&A,Combi2020}.
Furthermore, \cite{Marschall2020b} showed that the refractory-to-ice ratio is variable during the mission.
It was highest very early on in the mission ($\sim1.6$) when the value of \cite{Rotundi2015Sci} was measured.
Between perihelion and summer solstice, when the production rates were the highest and thus the time most representative of the bulk mass loss, the refractory-to-ice ratio dropped to $\sim0.5$.
The similarity between the values we find here and the ones measured in the coma \citep[][]{Biver2019A&A,Marschall2020b, Laeuter2020, Combi2020} suggests that the material falling back to the surface is not significantly different from that of the bulk comet, i.e., it did not lose a significant part of its ice component while in the coma.
This is in line with thermophysical models of fallback material, arguing for the retention of a significant amount of ices even for small cm-sized particles \citep{Davidsson2021Icar}.
Though we only discussed an upper limit for $\chi$ here, we will discuss a lower limit in the next section.

\subsection{The bulk composition of 67P}
\begin{figure}
	\includegraphics[width=\textwidth]{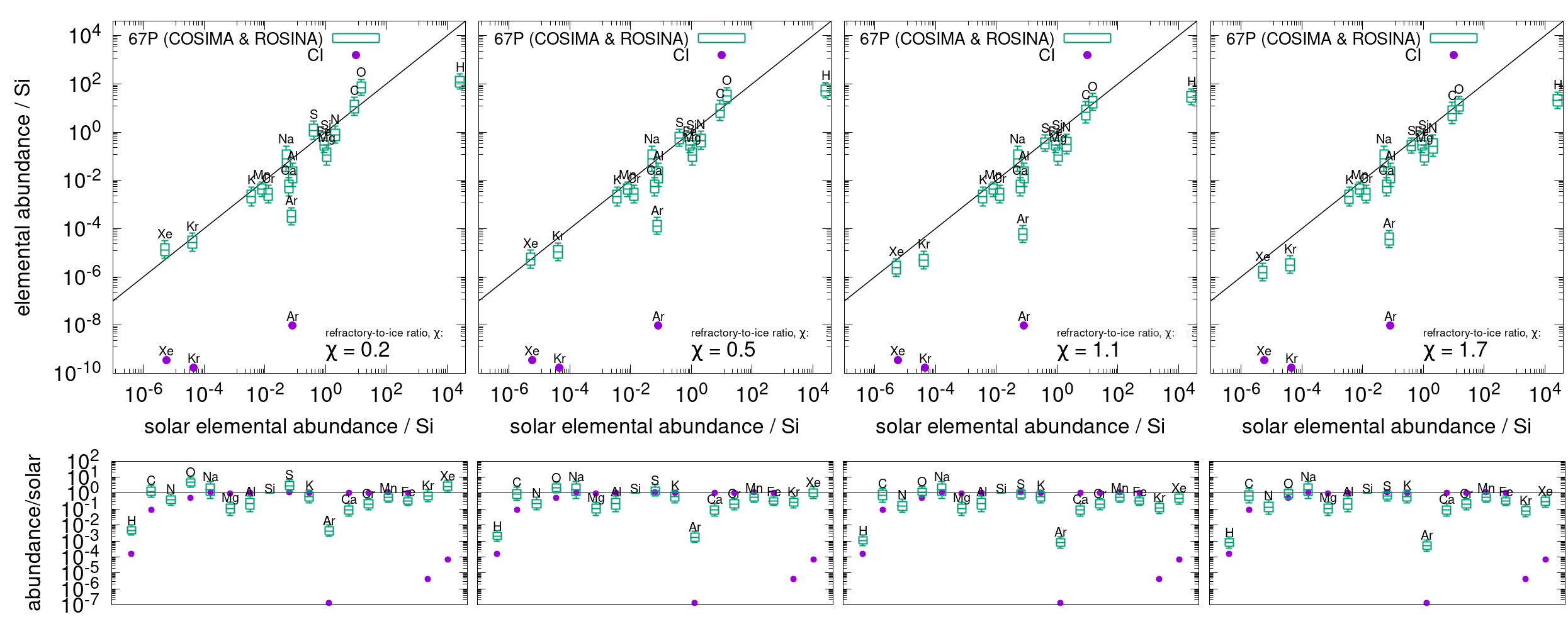}
	\caption{The top panels of each column show the bulk elemental composition of 67P with respect to silicon compared to the solar elemental composition \citep[][]{Asplund2021A&A}. Each column shows the results for a different refractory-to-ice mass ratio, $\chi$. The bottom panels of each column show the abundances of each element normalised to the solar values. The green boxes are the values for 67P, while the purple dots show the composition for CI chondrites \citep[][]{Alexander2019GeCoA}. All bulk abundances and corresponding errors for 67P are also available in Tables~\ref{tab:bulk67P1} and \ref{tab:bulk67P2}.}
	\label{fig:composition2Fe} 
\end{figure}

Given the arguments above, we have constrained the refractory-to-ice mass ratio to $\chi<1.7$.
Figure~\ref{fig:composition2Fe} presents an overview of the bulk abundances for 67P relative to the solar values for three different values of $\chi$.
The full list of values presented in that figure can be found in Tables~\ref{tab:bulk67P1} and \ref{tab:bulk67P2}.

Several things are of note. 
First, if $\chi<0.3$ xenon becomes super-solar in $84\%$ of the cases.
There is no particular reason that xenon should be super-solar, so we can consider this a mild lower limit on $\chi$.

Second, oxygen is super solar if $\chi<0.5$ and becomes consistent with the solar value at $\chi\sim0.5$.
Thus, a rather plausible range for $\chi$ is $0.5<\chi<1.7$.
For this range of $\chi$ the bulk mass fraction of water is between $28\%$ ($\chi=1.7$) to $52\%$ ($\chi=0.5$).
Interestingly, the prediction from \cite{Bitsch2020A&A} for a solar-type star is a water mass fraction of $35\%$.
This is well within our range with a water mass fraction of $35\%$ at $\chi=1.2$. 
Oxygen is solar in the median case at $\chi=1.1$ and thus represents our preferred case, which also happens to lie in the middle of our constraint of $\chi$ ($0.5<\chi<1.7$).
Similarly, \cite{Bitsch2020A&A} predict an oxygen mass fraction of $55\%$ whereas we get a range of $58\%$ ($\chi=1.7$) to $69\%$ ($\chi=0.5$).
An updated overview of literature values of $\chi$ for comet 67P can be found in the appendix in Fig.~\ref{fig:lit-ref-to-ice-ratio}.

Third, the noble gases argon, krypton, and xenon are all enriched by orders of magnitude compared to CI chondrites.

Fourth, and maybe most importantly, xenon (the least volatile of the three mentioned noble gases) is present in quasi-solar abundance for all the plausible cases of $\chi$ (within $1\sigma$ xenon is solar for $\chi<1.2$ and slightly sub-solar for larger $\chi$).
We can also look at Xe/H$_2$O for a sanity check, which comes only from the ROSINA measurements and does not rely on our modelling of the bulk abundances.
For 67P, Xe/H$_2$O~$=(2.4\pm1.1)\cdot10^{-7}$ \citep{Rubin2019MNRAS} while for the Sun using the values by \cite{Lodders2003ApJ} it is $(3.89\pm0.48)\cdot10^{-7}$ [assuming all oxygen is in water], $(6.95\pm0.86)\cdot10^{-7}$ [assuming $56\%$ of the oxygen atoms are in water as in the case of our $\chi=1.1$], and $(7.78\pm0.96)\cdot10^{-7}$ [assuming half of the oxygen is in water].
The 67P xenon to water ratio is sub-solar by factors of $1.6\pm0.8$, $2.9\pm1.4$, and $3.2\pm1.5$ for the three cases assuming different amounts of oxygen trapped in water.
Thus, even for the measurements only derived from ROSINA, xenon is quasi-solar.

Fifth, Krypton, which is slightly more volatile than xenon, is slightly depleted (factor of 3-10) with respect to the solar value.
Argon, on the other hand, the most volatile of the three noble gases, is strongly depleted ($\sim3$ orders of magnitude) with respect to the solar value.

\section{Discussion}
We focus our discussion on the elemental abundances of the noble gases (Fig.~\ref{fig:composition2Fe}).
Krypton is slightly depleted relative to xenon, and argon is heavily depleted relative to both krypton and xenon.
The important new constraint we introduce here is that xenon is present in the comet in solar abundance and about four orders of magnitude more abundant than in CI chondrites.

Noble gases can either be adsorbed in gaseous form in ice, or they can condense on icy grains if the disk temperature is low enough. 
Adsorption can only ever be partial. 
It cannot explain the solar abundance of xenon in 67P nor the fact that the isotopic properties of xenon in the comet are radically different from those of the Sun or the atmosphere of Jupiter \citep{Marty2017Sci}. 
Thus, we think that the data from 67P rule adsorption out for xenon.

Condensation of noble gases on grains can, in principle, explain solar abundance. 
Moreover, if at least part of the noble gases had been injected into the disk already in condensed form, they would have avoided isotopic equilibration with the gas and preserved a distinct isotopic composition. 
The condensation temperatures of noble gases on grains are given in \cite{Oberg2019AJ}: roughly 20~K for argon, 30~K for krypton, and 45~K for xenon. 
Explaining the depletion pattern in 67P, however, is not trivial. 
The reason is that once the original grains aggregate into pebbles, noble gases are not expected to sublimate at the corresponding temperatures. 
They remain trapped in the icy matrix and are desorbed only at specific temperatures, primarily during the transformation of CO$_2$ and H$_2$O ices \citep{Ligterink2024A&A}. 
These transformations start at about $50-60$~K and equally for all noble gases. 
Thus, in a scenario where grains drift towards the Sun from the outer disk, they would contain xenon, krypton and argon in solar abundance until they reach this temperature. 
According to this, any object formed in a region of the disk colder than $50-60$~K should have all noble gases in solar abundance, while objects formed in warmer portions of the disk can start fractionating their elemental ratios. 
Nevertheless, at any temperature, the strong depletion of argon relative to krypton and xenon observed in 67P is difficult to understand \citep{Ligterink2024A&A}, even more so considering our result that xenon is in solar abundance.

We propose to solve this conundrum taking into account that comets very likely formed late in the protoplanetary disk phase. 
In fact, comets and other related large icy small bodies such as Trojans and Kuiper-belt objects have very low densities \citep[between $300$~kg~m$^{-3}$ and $1500$~kg~m$^{-3}$;][]{Groussin2019,Berthier2020,Spencer2020} and contain highly-volatile ices, in particular CO and CO$_2$ \citep{Eberhardt1987A&A,Gasc2017MNRAS,Kelley2022ApJL}, which suggests that comets have never been significantly heated or even differentiated. Thus, they must have formed late \citep{Neumann2018JGRE} at a time when most of the short-lived radiogenic material has already decayed, i.e., several million years after the formation of the oldest solids \citep[CAIs;][]{Amelin2010E&PSL,Connelly2012Sci}.
At that time, dust should have been trapped in disk structures, rather than still drifting towards the star.
Evidence for this comes from the ubiquitous observation of dust rings in protoplanetary disks \citep{Andrews2020ARA&A} and the ratio between the disk sizes in dust and gas does not seem to shrink over time \citep{Najita2018ApJ}.
The observed evolution of disk sizes over time is totally inconsistent with models where dust continuously drifts towards the star \citep{Appelgren2020A&A, Birnstiel2024ARA&A}, demonstrating the need for dust traps due to radial pressure bumps. 
It is, therefore, reasonable to envision that dust remains trapped in pressure bumps for millions of years until the conditions (i.e., the dust/gas ratio, which increases as gas is removed) become favorable for late cometesimal formation.

\begin{figure}
	\includegraphics[width=\textwidth]{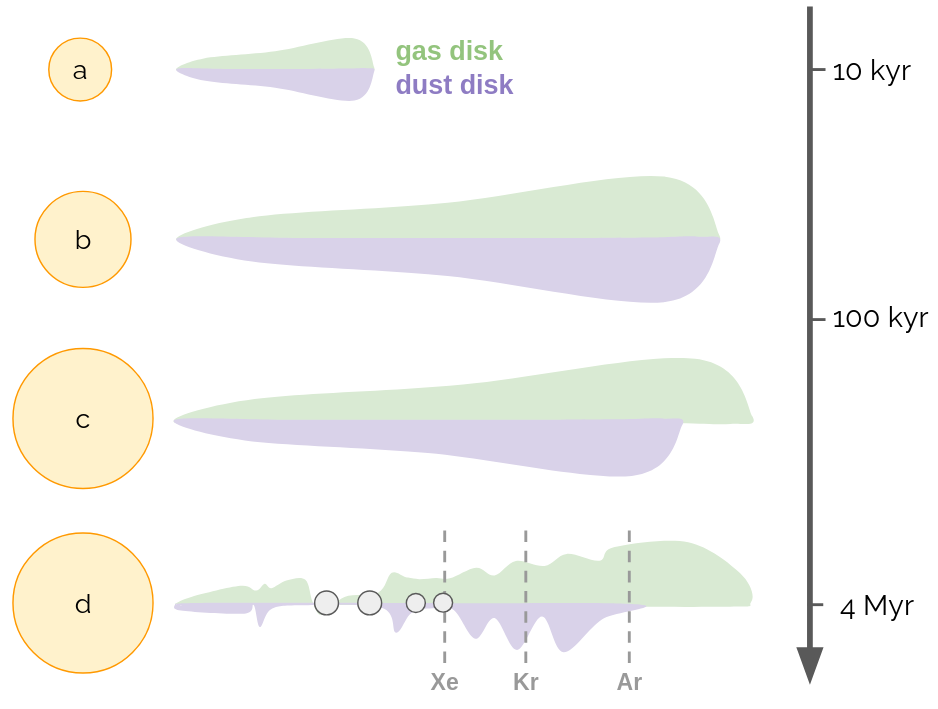}
	\caption{A sketch of a protoplanetary disk scenario. Initially, the dust and gas disk expand together through the viscous expansion of the gas \citep[e.g.,][]{Drazkowska2018A&A, Morbidelli2022NatAs, Marschall2023A&A} (a \& b). At some point, the dust grows large enough that radial inward drift due to aerodynamic drag in the tangential direction starts to dominate on short timescales \citep[e.g.,][]{Takeuchi2002ApJ, Takeuchi2005ApJ}, and the dust disk begins to contract (c). Eventually, structures in the disk form \citep{Andrews2020ARA&A} and trap dust, thus preventing the complete loss of the dust disk (d). Dust pebbles trapped within these pressure bumps remain there until they are accreted into planetesimals. Depending on the location in the disk and the accretion time, pebbles lose volatiles/retain the respective volatile species. The xenon, krypton, and argon lines are illustrated as examples.}
	\label{fig:sketch-disk} 
\end{figure}

Over these long timescales, the noble gases, even if initially trapped in water or CO$_2$ ice in the pebble, have plenty of time to diffuse to the surface of the pebble \citep{smith1997evidence, livingston2002general, Ligterink2024A&A}. 
Once the volatiles have reached the surface, they can desorb/sublimate if the disk temperature is high enough, i.e. larger than the condensation/sublimation temperatures of  \citep{Oberg2019AJ}. 
If a noble gas evaporates at the surface (e.g., argon), the diffusion continues, and over time the pebble becomes strongly depleted in that gas. 
If instead the temperature in the disk is too cold for evaporation of the noble gas at the surface of the pebble (e.g., xenon), diffusion stops and the pebble preserves its original abundance. 
Close but inwards of a sublimation line, we can envision that the process of diffusion and sublimation is slow, and in the end, the pebble is only partially depleted. 
This could be the case of krypton. 
This scenario is sketched in Fig.~\ref{fig:sketch-disk}.

\begin{figure}
	\includegraphics[width=\textwidth]{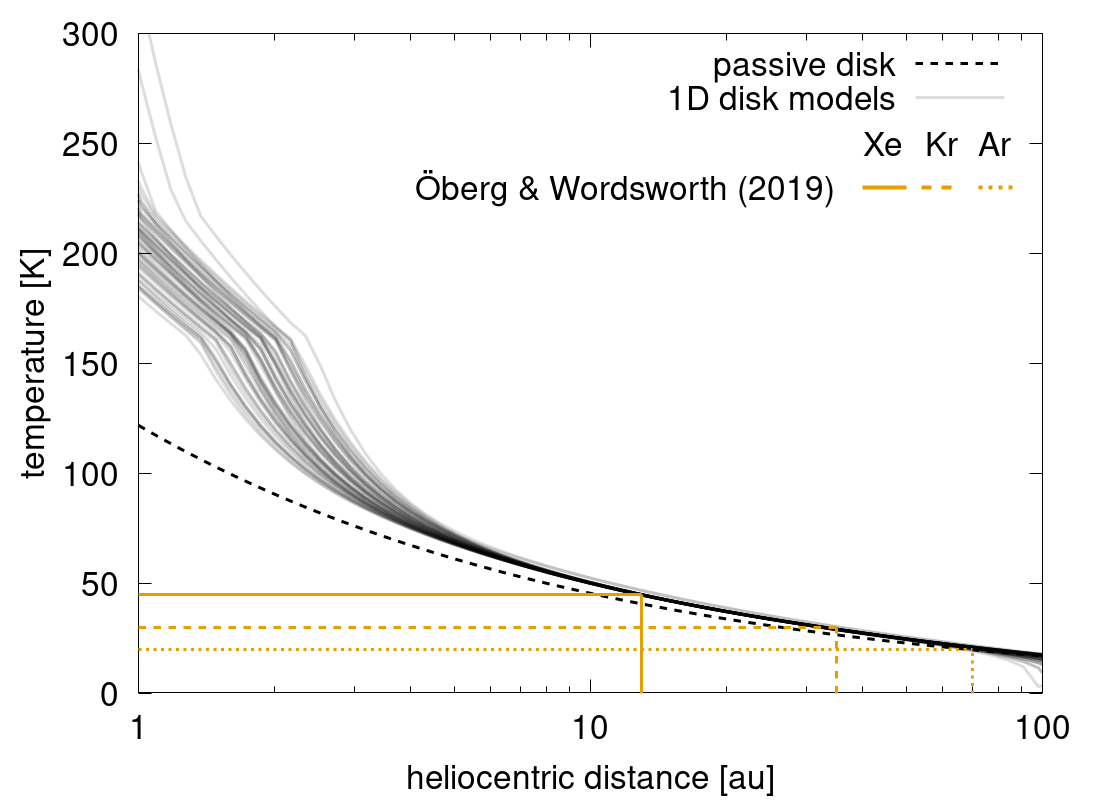}
	\caption{The temperature as a function of heliocentric distance is shown for a large set of models of evolved protoplanetary disks (light grey lines). The black dashed line shows the temperature profile of a passive disk \citep[][]{Chiang2010AREPS}. The orange lines show the sublimation lines for xenon (solid), krypton (dashed), and argon (dotted) for the values presented in \cite{Oberg2019AJ}. }
	\label{fig:disk-temperature} 
\end{figure}

Figure~\ref{fig:disk-temperature} shows the temperature profiles from our models of disks.
The parameters of the model cover the same values as in \cite{Marschall2023A&A} (e.g., in the assumed evolution of the disk viscosity).
Here, in contrast to the results presented in \cite{Marschall2023A&A}, we explicitly consider the stellar radiation onto the disk when solving for the temperature.
While the temperatures differ from model to model in the inner part of the disk, where viscous heating dominates, all of them show the same temperature profile beyond about 8~au.
Beyond that distance, the temperature is dominated by stellar radiation and thus follows almost exactly what would be expected from the analytical solution for a passive disk \citep[][]{Chiang2010AREPS}.
%We can, therefore, use these temperatures to see where the xenon, krypton, and argon lines of \cite{Oberg2019AJ} are in the disk.

The xenon line is at around 13~au, the krypton line is at roughly 45~au, and the one of argon is at 70~au.
We argued above that 67P must have formed beyond the xenon line, thus beyond 13~au, to accrete xenon in solid form.
67P should not form too close to the xenon line, though, because, in this case, it would form in a xenon-enriched environment that would lead to super-solar abundances of xenon.
Additionally, 67P also had to form close but still inside of the krypton line, i.e., inside of 35~au.
How far inside of the krypton line 67P could have formed is unclear.
But a reasonable guess would be that the formation region is somewhere in the 25-35~au region.
This conclusion, obtained from compositional considerations, is well in line with the location of the reservoir of planetesimals that generated the scattered disk, the Oort cloud and the hot Kuiper belt population, according to dynamical models \citep[][]{Nesvorny2018ARA&A}, and thus with the dynamical origin of 67P.

Additionally, CI chondrites are depleted relative to comets by at least two orders of magnitude even when we assume a $\chi=10$ (the order of magnitude for the amount of water in CIs; Fig.~\ref{fig:composition2Fe}).
In our scenario, this implies that CIs formed inside the xenon sublimation line, i.e., inside 12 au, namely the giant planet region.
This distance is also consistent with dynamical models of CI formation beyond Jupiter and implantation into the asteroid belt \citep{Raymond2017Icar}.

%\begin{figure}
%	\includegraphics[width=\textwidth]{figures/fig_noble-gases.png}
%	\caption{The isotopic composition of argon, krypton, and xenon are shown relative to the respective solar values \citep{Asplund2021A&A}. The purple points show the values for Ryugu \citep{Okazaki2023Sci}, the green points are for phase Q of chondrites \citep{Busemann2000M&PS}, and the yellow point are for the measurements by ROSINA for 67P of argon \citep{Balsiger2015SciAdv}, krypton \citep{Rubin2018SciA}, and xenon \citep{Marty2017Sci}. }
%	\label{fig:noble-gases} 
%\end{figure}

In our model, because the comet forms well inside the argon condensation line, the only argon present would be the one adsorbed from the protoplanetary disk. 
This is consistent with the results of \cite{Yokochi2022ApJ}, who showed experimentally that the adsorption of argon at 36-45~K (i.e. between the Kr and Xe lines) would lead to the observed Ar/H$_2$O ratio of 67P.
Additionally, argon, adsorbed from the gas of the disk, would have solar isotopic composition, as observed \citep{Balsiger2015SciAdv}.
%While both the elemental abundance and isotopic composition (Fig.~\ref{fig:noble-gases}) point to the accretion of argon from the gas phase, we cannot exclude that argon, and possibly krypton, was further depleted after formation. 
Both heating from solar radiation \citep[e.g.,][]{Guilbert-Lepoutre2016MNRAS,Gkotsinas2024PSJ} and collisions \citep[e.g.,][]{Bottke2023PSJ,Jutzi2020Icar} could further deplete argon.

\section{Conclusions}
We have combined COSIMA and ROSINA data on the elemental abundances of comet 67P's refractory and volatile/ice components.
To calculate the bulk elemental abundance of 67P, we started out being agnostic about the refractory-to-ice mass ratio, $\chi$, and left it as a free parameter.
The resulting elemental composition showed, though, that only $0.5<\chi<1.7$ leads to reasonable results.
These constraints on $\chi$ come from the amount of oxygen in water vs anhydrous material being not smaller than in CI chondrites as well as other elements (such as xenon and oxygen) not being in super-solar abundance relative to refractory elements.
These values are consistent with other estimates from the coma of 67P \citep[e.g.][]{Biver2019A&A,Marschall2020b, Laeuter2020, Combi2020, Choukroun2020SSRv} and suggest little volatile loss from material falling back to the nucleus, which is also consistent with thermophysical studies of the fallback material \citep{Davidsson2021Icar}.
For these values of $\chi$ comet 67P would have a $0.2< \text{C/O} <0.4$.
Furthermore, for the mentioned range in $\chi$, xenon is present in quasi-solar abundance. 
At the same time, krypton is slightly depleted relative to the solar value, and argon is heavily depleted.
This pattern naturally occurs if 67P formed from pebbles that were trapped and retained in the 25-35~au region for millions of years.
Over this retention time, diffusion causes the volatiles to reach the surface and sublimate if the temperature is high enough.
%By the time they reached this region, they would have lost their argon and part of their krypton to the gas disk but retained their xenon.
Xenon was thus accreted in the solid form and, therefore, in solar abundance.
In contrast, krypton was partially lost, and argon only accreted through adsorption from the gas.
%Xenon was also able to retain its pre-solar isotopic anomalies, whereas the ones potentially carried by argon and krypton would have been diluted in the gas phase by the solar isotopic composition.
%This is indeed what is observed in the ROSINA/Rosetta data.
%Finally, the fact that xenon shows signatures of r-process depletion suggests that a similar depletion should also characterise its refractory material. 
%Because different degrees of depletion in r-isotopes of refractory elements relative to solar (CAI) composition characterize the so-called isotopic dichotomy of NC and CC meteorites \citep[][]{Warren2011,Kruijer2017}, it becomes compelling to measure the isotopic composition of refractory material from comets to place them with respect to this dichotomy.
%Such measurements are achievable with a future no-cryogenic comet surface sample return because only the most refractory material would need to be measured.\\

\vspace{1cm}
{\noindent\large{Acknowledgements}}\\
We acknowledge the funding from the European Research Council (ERC) under the European Union’s Horizon 2020 research and innovation programme (Grant Agreement No. 101019380).
Additionally, we acknowledge support from programme ANR-20-CE49-0006 (ANR DISKBUILD).\\
We thank Martin Rubin for clarifying discussions about their work \citep{Rubin2019MNRAS} from which we used the elemental abundances derived from ROSINA data.
We thank Nicolas Fray, Ana\"is Bardyn, Hervé Cottin, and Martin Hilchenbach for similar clarifying discussions about the most up-to-date COSIMA data to use for the elemental composition of the refractories \citep{Bardyn2017MNRAS,Fray2017MNRAS}.
Furthermore, we thank Niels Ligterink for insightful discussions on ad-/desoption and diffusion of noble gases.


%\vspace{5mm}
%\facilities{}


%\software{%astropy \citep{2013A&A...558A..33A},  
%          }



\newpage
\appendix

\section{Elemental abundances of COSIMA and ROSINA}

\begin{figure}[h]
	\includegraphics[width=\textwidth]{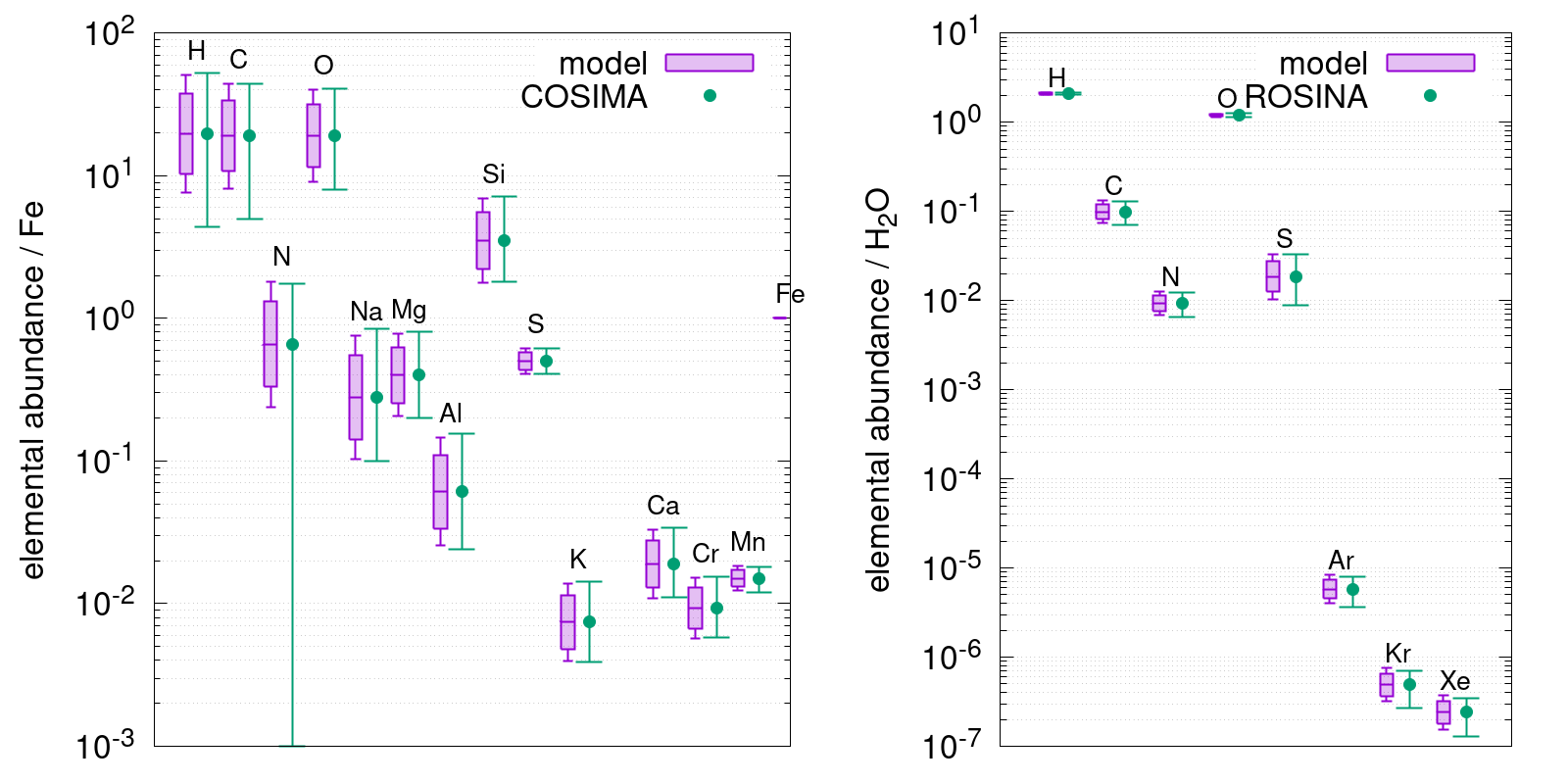}
	\caption{The left panel shows the elemental abundances (number of atoms) of the refractory component with respect to iron, Fe, as measured by COSIMA, while the right panel shows the elemental abundances (number of atoms) of the volatile/ice component with respect to molecular water, H$_2$O, as measured by ROSINA. The green points represent the measurements of the respective instruments, and the purple box plots the results from the one million random draws from the respective log-normal distributions. The upper and lower bounds of the box contain 50\% around the median while the whiskers contain 68\%, i.e. $\pm 1 \sigma$.}
	\label{fig:compare2ROSINAandCOSIMA} 
\end{figure}

Each element is assumed to follow a log-normal distribution 
\begin{equation*}
    \frac{1}{x\sigma\sqrt{2\pi}} \exp\left(-\frac{(\ln x - \mu)^2}{2\sigma^2} \right) \quad.
\end{equation*}
The values of $\mu$ and $\sigma$ are shown in Tables~\ref{tab:cosima-measurements} and \ref{tab:rosina-measurements}, and the comparison of the 'model' to the measurements is shown in Fig.~\ref{fig:compare2ROSINAandCOSIMA}.

\begin{table}[h]
\caption{The COSIMA elemental abundances \citep[][]{Bardyn2017MNRAS} with respect to Iron as well as the derived $\mu$ and $\sigma$ of the log-normal distribution used to model the data.}
\label{tab:cosima-measurements}
\begin{tabular}{llllllll}
\textbf{Z} & \textbf{Element} & \textbf{Name} & \textbf{X/Fe} & \textbf{-} & \textbf{+} & \textbf{$\mu$} & \textbf{$\sigma$} \\\hline
1	&	H	&	Hydrogen	&	19.76	&	15.36	&	33.04	&	2.98	&	0.96  \\
6	&	C	&	Carbon	&	19	&	14	&	25	&	2.94	&	0.85  \\
7	&	N	&	Nitrogen	&	0.66	&	0.69	&	1.1	&	-0.42	&	1.02  \\
8	&	O	&	Oxygen	&	19	&	11	&	22	&	2.94	&	0.75  \\
11	&	Na	&	Sodium	&	0.28	&	0.18	&	0.56	&	-1.27	&	1.01  \\
12	&	Mg	&	Magnesium	&	0.4	&	0.2	&	0.4	&	-0.92	&	0.67  \\
13	&	Al	&	Aluminum	&	0.061	&	0.037	&	0.095	&	-2.80	&	0.88  \\
14	&	Si	&	Silicon	&	3.5	&	1.7	&	3.7	&	1.25	&	0.68  \\
16	&	S	&	Sulfur	&	0.5	&	0.094	&	0.116	&	-0.69	&	0.21  \\
19	&	K	&	Potassium	&	0.0074	&	0.0035	&	0.0069	&	-4.91	&	0.63  \\
20	&	Ca	&	Calcium	&	0.019	&	0.008	&	0.015	&	-3.96	&	0.56  \\
24	&	Cr	&	Chromium	&	0.0093	&	0.0035	&	0.0062	&	-4.68	&	0.49  \\
25	&	Mn	&	Manganese	&	0.015	&	0.003	&	0.003	&	-4.20	&	0.20  \\
26	&	Fe	&	Iron	&	1	&	0	&	0	&	0.00	&	0.00  \\
\end{tabular}
\end{table}

\begin{table}[h]
\caption{The ROSINA elemental abundances \citep[][]{Rubin2019MNRAS} with respect to water as well as the derived $\mu$ and $\sigma$ of the log-normal distribution used to model the data.}
\label{tab:rosina-measurements}
\begin{tabular}{llllllll}
\textbf{Z} & \textbf{Element} & \textbf{Name} & \textbf{X/H$_2$O} & \textbf{-} & \textbf{+} & \textbf{$\mu$} & \textbf{$\sigma$} \\
1  & H  & Hydrogen & 2.1002E+00 & 3.1708E-02 & 3.4491E-02 & 0.74   & 0.02 \\
6 & C & Carbon & 9.8983E-02 & 2.9185E-02 & 3.0450E-02 & -2.31  & 0.29 \\
7 & N & Nitrogen & 9.3630E-03 & 2.8593E-03 & 2.8593E-03 & -4.67  & 0.30 \\
8 & O & Oxygen & 1.1980E+00 & 6.3501E-02 & 6.9131E-02 & 0.18   & 0.06 \\
16 & S & Sulfur & 1.8557E-02 & 9.7949E-03 & 1.4148E-02 & -3.99  & 0.59 \\
18 & Ar & Argon & 5.8000E-06 & 2.2000E-06 & 2.2000E-06 & -12.06 & 0.37 \\
36 & Kr & Krypton & 4.9000E-07 & 2.2000E-07 & 2.2000E-07 & -14.53 & 0.43 \\
54 & Xe & Xenon & 2.4000E-07 & 1.1000E-07 & 1.1000E-07 & -15.24 & 0.44 
\end{tabular}
\end{table}

\begin{figure}
	\includegraphics[width=\textwidth]{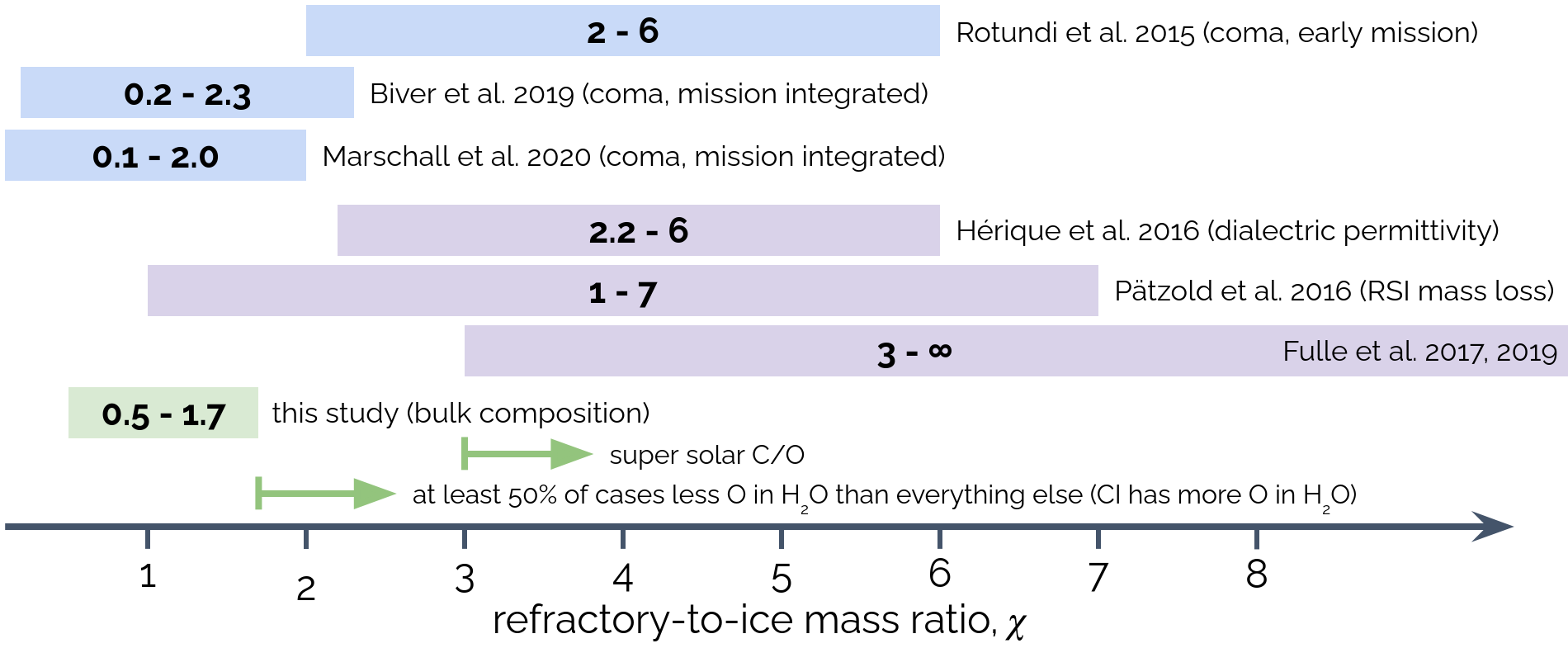}
	\caption{Summary of literature values for the refractory-to-ice mass ratio, $\chi$: Measurements in the coma are shown in blue and include the ones derived early in the mission from the GIADA instrument \citep{Rotundi2015Sci}, integrated values combining coma gas/dust measurements and the total mass loss \citep{Biver2019A&A, Marschall2020b}. Estimates from nucleus measurements are shown in purple \citep{Herique2016MNRAS,Patzold2016Natur,Fulle2017MNRAS,Fulle2019MNRAS}. The value we constrain here purely from the elemental composition is shown in green. Additionally, we show the values of $\chi$ for which the C/O ratio becomes super solar and the value of $\chi$ when at least 50\% of cases have less oxygen in water than anything else. }
	\label{fig:lit-ref-to-ice-ratio} 
\end{figure}

\newpage
\section{Bulk elemental abundances of 67P}

\begin{table}[h]
\caption{Bulk elemental abundances of 67P with respect to silicon for the refractory-to-ice mass ratio, $\chi=0.3$ and $0.5$. The errors are $\pm 1\sigma$.}
\label{tab:bulk67P1}
\begin{tabular}{lll|lll|lll|}
\textbf{}  & \textbf{}        & \textbf{}     & \textbf{$\chi=0.3$} & \textbf{}  & \textbf{}  & \textbf{$\chi=0.5$} & \textbf{}  & \textbf{}  \\
\textbf{Z} & \textbf{element} & \textbf{name} & \textbf{X/Si}       & \textbf{+} & \textbf{-} & \textbf{X/Si}       & \textbf{+} & \textbf{-} \\ \hline
1          & H                & Hydrogen      & 8.2689E+01          & 9.5216E+01 & 4.1470E+01 & 5.2849E+01          & 6.1041E+01 & 2.6713E+01 \\
6          & C                & Carbon        & 9.2908E+00          & 1.4429E+01 & 5.3951E+00 & 7.7621E+00          & 1.2871E+01 & 4.6455E+00 \\
7          & N                & Nitrogen      & 5.8488E-01          & 7.9705E-01 & 3.1997E-01 & 4.3520E-01          & 6.4395E-01 & 2.4500E-01 \\
8          & O                & Oxygen        & 4.8396E+01          & 5.8151E+01 & 2.4576E+01 & 3.1377E+01          & 3.8295E+01 & 1.6105E+01 \\
11         & Na               & Sodium        & 7.9736E-02          & 1.8869E-01 & 5.6142E-02 & 7.9736E-02          & 1.8869E-01 & 5.6142E-02 \\
12         & Mg               & Magnesium     & 1.1364E-01          & 1.8106E-01 & 6.9558E-02 & 1.1364E-01          & 1.8106E-01 & 6.9558E-02 \\
13         & Al               & Aluminum      & 1.7209E-02          & 3.4748E-02 & 1.1480E-02 & 1.7209E-02          & 3.4748E-02 & 1.1480E-02 \\
14         & Si               & Silicon       & 1.0000E+00          & 0.0000E+00 & 0.0000E+00 & 1.0000E+00          & 0.0000E+00 & 0.0000E+00 \\
16         & S                & Sulfur        & 8.2997E-01          & 1.1444E+00 & 4.5791E-01 & 5.6403E-01          & 7.3344E-01 & 3.0396E-01 \\
19         & K                & Potassium     & 2.1090E-03          & 3.1999E-03 & 1.2671E-03 & 2.1090E-03          & 3.1999E-03 & 1.2671E-03 \\
18         & Ar               & Argon         & 2.0555E-04          & 2.7418E-04 & 1.1058E-04 & 1.2333E-04          & 1.6451E-04 & 6.6349E-05 \\
20         & Ca               & Calcium       & 5.4035E-03          & 7.6795E-03 & 3.1547E-03 & 5.4035E-03          & 7.6795E-03 & 3.1547E-03 \\
24         & Cr               & Chromium      & 2.6481E-03          & 3.4905E-03 & 1.5036E-03 & 2.6481E-03          & 3.4905E-03 & 1.5036E-03 \\
25         & Mn               & Manganese     & 4.2759E-03          & 4.3790E-03 & 2.1680E-03 & 4.2759E-03          & 4.3790E-03 & 2.1680E-03 \\
26         & Fe               & Iron          & 2.8590E-01          & 2.7567E-01 & 1.4117E-01 & 2.8590E-01          & 2.7567E-01 & 1.4117E-01 \\
36         & Kr               & Krypton       & 1.7435E-05          & 2.4472E-05 & 9.6898E-06 & 1.0461E-05          & 1.4683E-05 & 5.8139E-06 \\
54         & Xe               & Xenon         & 8.5395E-06          & 1.1927E-05 & 4.7560E-06 & 5.1237E-06          & 7.1563E-06 & 2.8536E-06
\end{tabular}
\end{table}

\begin{table}[h]
\caption{Continuation of Table~\ref{tab:bulk67P1} with the bulk elemental abundances of 67P with respect to silicon for the refractory-to-ice mass ratio, $\chi$, of 1.1 and 1.7. The errors are $\pm 1\sigma$.}
\label{tab:bulk67P2}
\begin{tabular}{lll|lll|lll|}
\textbf{}  & \textbf{}        & \textbf{}     & \textbf{$\chi=1.1$} & \textbf{}  & \textbf{}  & \textbf{$\chi=1.7$} & \textbf{}  & \textbf{}  \\
\textbf{Z} & \textbf{element} & \textbf{name} & \textbf{X/Si}       & \textbf{+} & \textbf{-} & \textbf{X/Si}       & \textbf{+} & \textbf{-} \\ \hline
1          & H                & Hydrogen      & 2.7996E+01          & 3.3598E+01 & 1.4465E+01 & 2.0543E+01          & 2.5603E+01 & 1.0819E+01 \\
6          & C                & Carbon        & 6.4843E+00          & 1.1635E+01 & 4.0489E+00 & 6.1019E+00          & 1.1317E+01 & 3.8755E+00 \\
7          & N                & Nitrogen      & 3.0525E-01          & 5.2457E-01 & 1.8182E-01 & 2.6449E-01          & 4.9492E-01 & 1.6259E-01 \\
8          & O                & Oxygen        & 1.7401E+01          & 2.2356E+01 & 9.1901E+00 & 1.3257E+01          & 1.7670E+01 & 7.1568E+00 \\
11         & Na               & Sodium        & 7.9736E-02          & 1.8869E-01 & 5.6142E-02 & 7.9736E-02          & 1.8869E-01 & 5.6142E-02 \\
12         & Mg               & Magnesium     & 1.1364E-01          & 1.8106E-01 & 6.9558E-02 & 1.1364E-01          & 1.8106E-01 & 6.9558E-02 \\
13         & Al               & Aluminum      & 1.7209E-02          & 3.4748E-02 & 1.1480E-02 & 1.7209E-02          & 3.4748E-02 & 1.1480E-02 \\
14         & Si               & Silicon       & 1.0000E+00          & 0.0000E+00 & 0.0000E+00 & 1.0000E+00          & 0.0000E+00 & 0.0000E+00 \\
16         & S                & Sulfur        & 3.4247E-01          & 4.0175E-01 & 1.7770E-01 & 2.7567E-01          & 3.0737E-01 & 1.4067E-01 \\
19         & K                & Potassium     & 2.1090E-03          & 3.1999E-03 & 1.2671E-03 & 2.1090E-03          & 3.1999E-03 & 1.2671E-03 \\
18         & Ar               & Argon         & 5.6059E-05          & 7.4775E-05 & 3.0159E-05 & 3.6273E-05          & 4.8384E-05 & 1.9514E-05 \\
20         & Ca               & Calcium       & 5.4035E-03          & 7.6795E-03 & 3.1547E-03 & 5.4035E-03          & 7.6795E-03 & 3.1547E-03 \\
24         & Cr               & Chromium      & 2.6481E-03          & 3.4905E-03 & 1.5036E-03 & 2.6481E-03          & 3.4905E-03 & 1.5036E-03 \\
25         & Mn               & Manganese     & 4.2759E-03          & 4.3790E-03 & 2.1680E-03 & 4.2759E-03          & 4.3790E-03 & 2.1680E-03 \\
26         & Fe               & Iron          & 2.8590E-01          & 2.7567E-01 & 1.4117E-01 & 2.8590E-01          & 2.7567E-01 & 1.4117E-01 \\
36         & Kr               & Krypton       & 4.7550E-06          & 6.6742E-06 & 2.6427E-06 & 3.0767E-06          & 4.3186E-06 & 1.7100E-06 \\
54         & Xe               & Xenon         & 2.3289E-06          & 3.2529E-06 & 1.2971E-06 & 1.5070E-06          & 2.1048E-06 & 8.3929E-07
\end{tabular}
\end{table}

\newpage
\section{Using CI composition for the refractory component}
One might imagine that instead of the COSIMA measurements, one could use the composition of CI chondrites for the refractory component.
If one were to do that, how would our results change?

\begin{figure}
	\includegraphics[width=\textwidth]{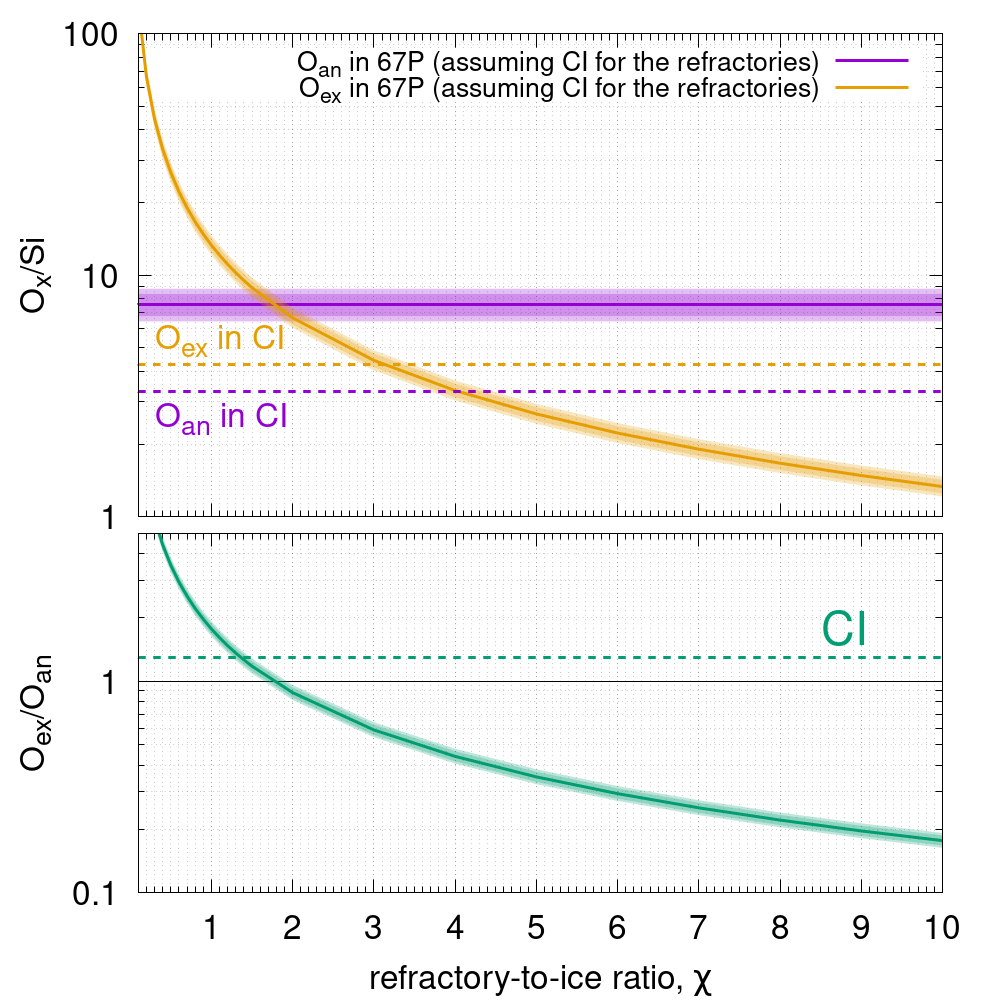}
	\caption{The top panel shows the oxygen-to-silicon ratio for oxygen that is in water (subscript 'ex') and that which is in anhydrous form (subscript 'an'). Compared to Fig~\ref{fig:oxygen} we assume CI composition for the refractory component of 67P. The values for CI chondrites are taken from \cite{Alexander2019GeCoA}. The bottom panel shows the ratio of the two oxygen components. Note that the bottom panel is not simply the ratio of the lines in the top panel. O$_{ex}$ and O$_{an}$ are not uncorrelated, and therefore, their ratio and respective errors need to be computed for each random draw. The darker shaded areas contain 50\% of the cases while the lighter areas contain $\pm 1 \sigma$}
	\label{fig:oxygenCI} 
\end{figure}

If we replace the values used from COSIMA (Table~\ref{tab:cosima-measurements}) with CI composition, the constraint on the refractory-to-ice ratio becomes $0.35 < \chi < 1.35$, which is very similar to what we find using the COSIMA composition ($0.5 < \chi < 1.7$).
Figure~\ref{fig:oxygenCI} is equivalent to Fig.~\ref{fig:oxygen} that we used to constrain the upper limit for $\chi$.
For this range ($0.35 < \chi < 1.35$), xenon is solar when $0.35 < \chi < 0.9$.
At the upper bound, when $\chi = 1.35$, xenon is slightly sub-solar by 30\%.
Therefore, even if comet 67P were composed of refractories with CI composition, all our conclusions hold.

\newpage

\bibliography{main}{}
\bibliographystyle{aasjournal}

%\bibliographystyle{alpha}
%\bibliography{sample}

\end{document}